\documentclass[aps,prl,twocolumn,showpacs,byrevtex,superscriptaddress]{revtex4}
\usepackage{epsfig}

\usepackage{epsfig}

\def\be{\begin{equation}}
\def\eea{\end{eqnarray}}
\def\ee{\end{equation}}
\def\bea{\begin{eqnarray}}
\def\ea{\end{array}}
\def\ba{\begin{array}}

\newcommand{\exval}[1]{\mbox{$\langle \, {#1}\, \rangle$}}
\newcommand{\bel}[1]{\begin{equation}\label{#1}}

\def\zzz{{\mathchoice {\hbox{$\sf\textstyle Z\kern-0.4em Z$}}
{\hbox{$\sf\scriptstyle Z\kern-0.3em Z$}}
{\hbox{$\sf\scriptscriptstyle Z\kern-0.2em Z$}}
{\hbox{$\sf\textstyle Z\kern-0.4em Z$}}}}

\begin{document}

\title{
Amnestically
induced persistence in random walks}

\author{J. C. Cressoni} 
\affiliation{\mbox{Instituto de F\'{\i}sica, Universidade Federal de
Alagoas,} Macei\'{o}--AL, 57072-970, Brazil}

\author{Marco Antonio Alves da Silva}
\affiliation{Departamento de F\'{\i}sica e Qu\'{\i}mica, FCFRP,
Universidade de S\~ao Paulo, 14040-903 Ribeir\~ao Preto, SP, Brazil}

\author{G. M. Viswanathan}
\affiliation{\mbox{Instituto de F\'{\i}sica, Universidade Federal de
Alagoas,} Macei\'{o}--AL, 57072-970, Brazil}


\revised{\today}

\vspace{2cm}

\begin{abstract}
We study how the Hurst exponent $\alpha$ depends on the fraction $f$
of the total time $t$ remembered by non-Markovian random walkers that
 {recall} only the distant past. We find that otherwise
nonpersistent random walkers switch to persistent behavior when
inflicted with significant memory loss.   Such memory losses induce the
probability density function of the walker's position to undergo a
transition from Gaussian to non-Gaussian.
%
We interpret these findings of
persistence in terms of a breakdown of self-regulation mechanisms and
discuss their possible relevance to some of the burdensome behavioral
and psychological symptoms of Alzheimer's disease and other
dementias.
\end{abstract}


\maketitle

A classic problem in physics concerns normal versus anomalous
diffusion~\cite{bachelier,shlesinger-book,metzlerklafterrev2000,metzlerklafterrev2004}.
Fractal analysis~\cite{bbm,bunde-havlin} of random
walks~\cite{bachelier,shlesinger-book,metzlerklafterrev2000,metzlerklafterrev2004}
with memory aims at quantitatively describing the complex
phenomenology observed in economic~\cite{bachelier,stanley-econo},
sociological~\cite{westbooksoc},
ecological~\cite{turchinbook,nature1996,nature1999},
biological~\cite{shlesinger-book,maa4},
physiological~\cite{shlesinger-book,ary} and
physical~\cite{bunde-havlin,shlesinger-book,metzlerklafterrev2000,metzlerklafterrev2004,herrmann}
systems. Markov processes exhaustively account for random walks with
short-range memory. In contrast, long-range memory typically gives
rise to non-Markovian
walks~\cite{bunde-havlin,shlesinger-book,stanley-econo,elefante,scv,hurst,sornette-pnas}.
The most extreme case of a non-Markovian random walk corresponds to a
stochastic process with dependence on the entire history of the
system.  With the aim of capturing the essential dynamics of memory
loss in complex systems, we investigate an idealized model for the
limiting case of unbounded memory random walks with dependence on the
complete~\cite{elefante} or partial~\cite{scv} history of a binary
decision process.
Persistent random walkers tend to repeat past behavior, hence a
 plausible {assumption} holds that loss of memory of the past cannot
 cause persistence but rather can only diminish it.
Moreover, all Markovian and most non-Markovian random walk models that
attempt to account for persistent {behavior} inevitably assume a
memory of the recent past, with memory loss limited to the distant
past.  Aiming to question such assumptions, we show here that loss of
memory of the recent rather than of the distant past can actually
induce persistence.

An important global property of random walks, the Hurst exponent
$\alpha$~\cite{hurst}, relates to how persistently the walkers diffuse.
For the case of zero drift velocity, the Hurst exponent quantifies how
the mean squared displacement scales with time $t$:
\bel{x2} \exval{x^2}\sim t^ {2\alpha}\;\;.
\ee
{The dynamics of the random walker can 
range from subdiffusion ($\alpha<1/2$), through normal diffusion
($\alpha = 1/2$) to superdiffusion ($\alpha>1/2$),  the latter
characterized by persistence (i.e., 
long-range correlations) in the random walk. Persistent random walkers
on average repeat past behavior.}

We first describe the case without memory loss~\cite{elefante,scv}. The
random walk starts at the origin at time $t_0=0$ and retains memory of
its complete history.
At each time step the random walker moves either one step to the right
or left:
$
x_{t+1} = x_t + v_{t+1}
$ 
where the velocity $v_{t+1} = \pm 1$ represents a stochastic noise
with two-point autocorrelations (i.e, memory).  The walker can
remember the entire history of prior random walk step directions
$\{v_{t'}\}$ for $t' \leq t$. 
At time $t$, we randomly choose a previous time $1\leq t'<t$ with
equal {\it a priori} probabilities.
We then choose the current step direction $v_{t}$ based on the
value of $v_{t'}$ in the following manner: 
\bel{eq-sigma}
v_t = \left\{ 
\begin{array}{ll} 
+v_{t^\prime} &\mbox{with probability $p$}\\
-v_{t^\prime} &\mbox{with probability $1-p$}
\end{array} \right.
\ee
Without loss of generality we assume that the first step always goes
to the right, i.e. $v_1=+1$.
The position at time $t$ thus follows
$x_t = \sum_{t'=1}^t v_{t'}\; \;$ .

\begin{figure}
\centerline{\psfig{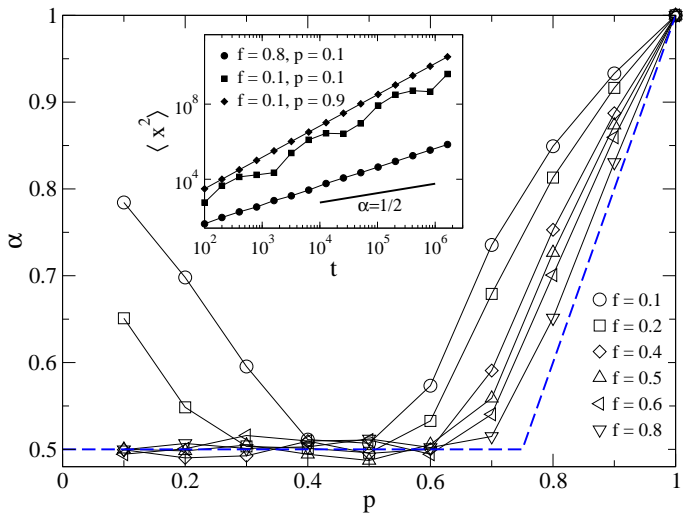}}
\caption{Persistence of the random walk, represented by the Hurst
exponent $\alpha$ as a function of the correlation parameter $p$ and
the fraction $f$ of the total time remembered by non-Markovian walkers
that forget the most recent $(1-f)t$ time steps, measured over $10^3$
realizations.  The dashed line shows the known analytic
result\protect~\cite{elefante,scv} for the case $f=1$. The inset shows
the typical plots of $\exval{x^2}$ for chosen values of $p$ and $f$
from which we estimated the $\alpha$. We find that whereas values
$p<1/2$ always preclude persistent behavior in the $f=1$ case, yet
persistence arises for sufficiently small $f$ even for $p<1/2$. Note
how the $f=0.2$ and $f=0.1$ curves deviate away from $\alpha=0.5$ as
$p\rightarrow 0$. We find that this persistence emerges as a result of
log-periodic oscillations (e.g., see the $f=p=0.1$ case in the inset)
in the velocity, such that it changes sign increasingly infrequently.
Note also that loss of recent memories increases persistence for the
entire range of $p\neq 1/2$.}
\label{fig-alpha}
\end{figure}

\begin{figure}
\centerline{\psfig{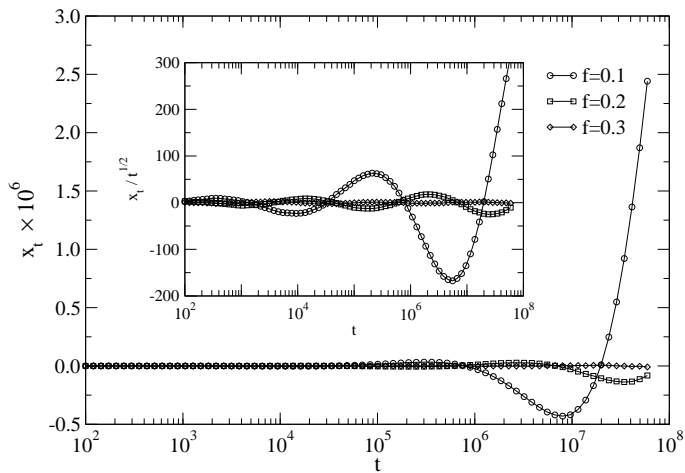}}
\caption{Semilog plot of displacement $x_t$ as a function of $t$ for
$p=0.1$ and various $f.$ The inset shows $x_t/t^{1/2}$ versus
time. Significant memory loss (i.e., small $f$) leads to larger
amplitudes for the log-periodic oscillations.}
\label{fig-logp}
\end{figure}

\begin{figure}
\centerline{\psfig{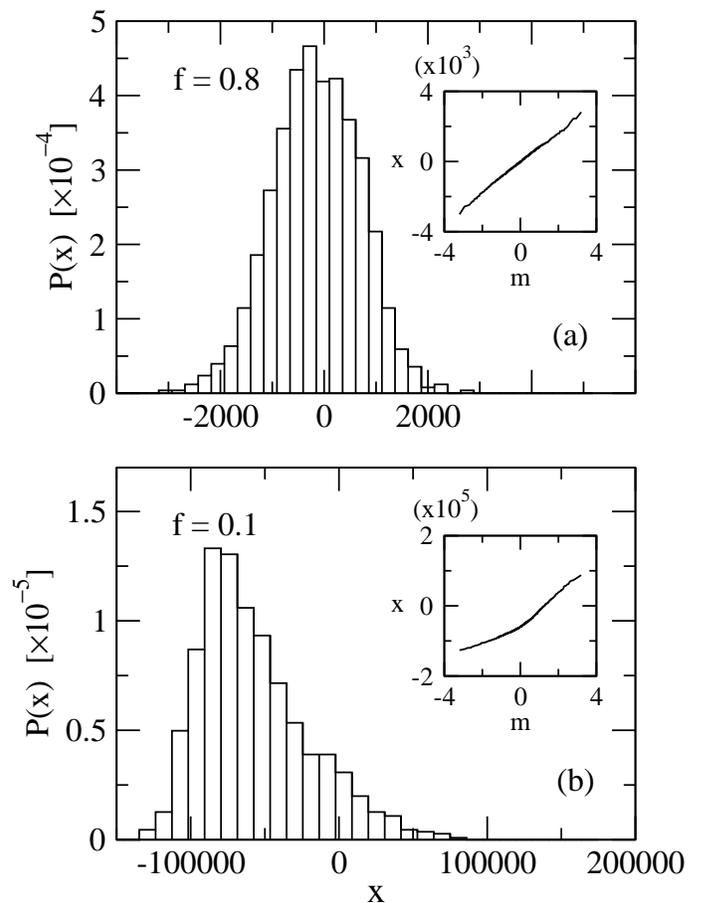}}
\caption{ Histogram showing the normalized probability density
function (PDF) of the position of walkers with $p=0.1$ after a long
time $t_{max}=$1638400, for (a) $f=0.8$ (nonpersistent regime) and (b)
$f=0.1$ (amnestically induced persistence).  We have simulated $10^3$
realizations. Notice the Gaussian PDF for large $f$, as expected from
the known properties of the $f=1$ case.  In contrast, the PDF becomes
non-Gaussian for low $f$ and low $p$, when memory loss induces
persistence. Indeed, the
aforementioned
log-periodic velocity inversions (see text) allow a
violation of the necessary conditions for the central limit theorem to
hold, thus preventing the convergence to a Gaussian PDF.  The insets
show normal probability plots of the position $x$ versus the normal
order statistic medians $m$: the linear plot in (a) indicates Gaussian
(i.e., normal) behavior, whereas the nonlinearity in (b) indicates a
non-Gaussian PDF.  The loss of recent memories leads to a remarkable
change from Gaussian to non-Gaussian behavior---and consequently to
persistence.}
\label{fig-pdf}
\end{figure}

The advantage of this choice of non-Markovian random walk model stems
from its known exact analytical solution~\cite{elefante}. The
probability density function (PDF) evolves according to a Gaussian
propagator with a diffusion constant that depends on time $t$ and $p$:
\bea \label{eq-P} P(x,t) &=& \frac{1}{\sqrt{4\pi tD(t)}}
\exp{\left[-\frac{\big(x-\exval{x(t)}\big)^2}{4tD(t)}\right]} \\
\label{eq-D} 
D(t,p) &=& \frac{1}{8p-6} \left[
\left(\frac{t}{t_0}\right)^{4p-3}-1 \right] \;\;.
\eea 
{Asymptotically, the model presents nonpersistent behavior
($\alpha=1/2$) for $p<3/4$ and a persistence regime ($\alpha=2p-1$)
for $p>3/4$ (with marginal persistence for
$p=3/4$)~\cite{elefante,scv}.
}
The mean displacement scales as $\exval{x}\sim t^{2p-1}$, decaying
algebraically for $p<1/2$
and diverging algebraically for $p>1/2$.  For $1/2<p<3/4$, the mean
square displacement remains larger than the square of the mean, such
that the behavior remains diffusive, i.e. nonpersistent. 
Besides the classification of the second moment behavior in terms of
nonpersistent (i.e., normal) versus persistent (i.e., anomalous)
diffusion (defined by the transition at $p=3/4$),
the first moment scaling allows the classification of the random
walkers as either ``reformers'' ($p<1/2$) that attempt to compensate
for the past behavior, or as ``traditionalists'' ($p>1/2$) that tend
to repeat the past.
Crucially, reformers ($p<1/2$) never show persistence
\mbox{($\alpha>1/2$).}  We next modify the model {in order to}
introduce loss of memory of the recent past.

Consider a random walker that can remember only the initial fraction
$ft$ of the total $t$ time steps. If $f=1$ then we recover the full
memory model, but for $f<1$ the walker, while remaining non-Markovian,
nevertheless does not remember the complete history.
We study how $\exval{x^2}$ scales with $t$, as a function of $p$ and
$f$.  Fig.~\ref{fig-alpha} shows how the scaling exponent $\alpha$ of
the mean square displacement varies as we reduce the range of the
memory.  
As $f\rightarrow 0$ we obtain an unexpected result: even reformers
($p<1/2$) that tend to compensate for past behavior become persistent
($\alpha>1/2$). The loss of memory of the recent past thus appears to
cause persistence for values of $p$ for which the full memory
model precludes persistence.  In contrast, for loss of recall of the
distant rather than recent past, the persistence can only
decrease~\cite{scv}.

How can loss of memory lead to persistence?
Note that for small $f$ and $p<1/2$, the random walker attempts to
move opposite  to the average direction chosen in the first $ft$ time
steps.  It takes $t(1-f)/f$
additional time steps for the effect of the action taken at the
present time $t$ to enter into the range of the accessible memory.
Therefore considerable time elapses between inversions of the time
averaged velocity.  The behavior thus becomes persistent, because the
mean position oscillates with ever greater amplitude as $t\rightarrow
\infty$, due to the
increasingly infrequent velocity inversions.  We have found evidence
of log-periodic oscillations, indicating that discrete scale
invariance, characterized by complex rather than real scaling
exponents, plays a role~\cite{sornette-pnas}.  Fig.~\ref{fig-logp}
shows clear evidence of the existence of these log-periodic
oscillations.  The amplitude of these oscillations becomes larger for
small $f$, whereas for sufficiently large $f$ they effectively
disappear.  In the future, we expect to obtain the complete phase
diagram separating the persistent and nonpersistent phases as a
function of $p$ and $f$.

We also study a more dramatic loss of memory---the random walkers'
equivalent of anterograde amnesia:  no new long-term memories can form
after a simulated ``accident'' or ``injury'' at time 
$t_a$.
In this case, we find ballistic ($\alpha=1$) behavior, with no
velocity inversions (or log-periodic oscillations) for any value
$p\neq1/2$.

We next report evidence that this amnestically induced persistence for
$p<1/2$ is associated with a transition from a Gaussian PDF of the
position for $f=1$ to a non-Gaussian PDF for $f<1$
(Fig.~\ref{fig-pdf}).  The inset in Fig.~\ref{fig-pdf} shows a normal
probability plot~\cite{pp} of the position.
On the vertical axis we plot the ordered data, and on the horizontal
axis we plot the normal order statistic medians for the normal (i.e.,
Gaussian) distribution. Departures from a linear plot indicate
departures from Gaussian statistics. We obtain a good fit with a
Gaussian distribution for large $f$ but not for small $f$.  One would
not typically expect this result, since only Gaussian PDFs have
appeared in similar models~\cite{elefante,scv}.  By changing the
parameter $f$, the behavior undergoes a remarkable qualitative change,
from Gaussian to non-Gaussian.

These numerical results are consistent with the known analytical
solutions for $f=1$~\cite{elefante} and for the case of random walkers
with anterograde amnesia, which leads to a ballistic solution with
constant velocity for $p\neq1/2$.  Analytic solutions for $0<f<1$ may
require solving recurrence relations in which the moments
$\exval{x^q}$ at time $t+1$ depend not only on their values at a
previous time $t$ but also on the values at a time $ft$ in the more
distant past.  The full analytical solution for $\alpha$ as a function
of $f$ and $p$ remains an open problem. In this context, we have found
preliminary evidence that the recurrence relations allow power law and
log-periodic solutions in the first moment $\exval{x_t}$.

We also discuss the expected behavior for $d$-dimensional
generalizations for $f<1$. For the case of separate memories for each
space direction, we expect the dimensionality not to play a major role
on how $\alpha$ depends on $f$.  The velocities now become
$d$-dimensional vectors, as do the displacements. But the velocities
along a given dimension do not affect the position along a distinct
dimension, thereby effectively decoupling the dimensions, in
accordance with the known analytical result for the $f=1$
model~\cite{elefante}.

We briefly comment on the above results, in the context of complex
systems that have learning or self-regulating mechanisms~\cite{maa3}
that preempt repetitive dynamics. By incorporating some form of
negative feedback, many complex adaptive systems avoid persistent or
repetitive behavior.  For the full memory model ($f=1$), negative
feedback occurs for $p<1/2$, so persistence cannot arise. However, for
$f<1$ the velocity inversions provoked by the negative feedback happen
ever more infrequently for the reasons explained earlier, with
important consequences for self-regulation.  Essentially, negative
feedback breaks down with the loss of recent recall, thus allowing
persistence.  Our results suggest the possibility of a new
quantitative
description of the phenomenology of memory loss and may achieve a
closer connection with realistic applications. Specifically, we note
(i) the possibility of quantifying the extent of memory damage in
diverse (e.g., neurophysiological) non-Markovian systems via $1-f$ and
(ii) the known relationship between $\alpha$ and
pathology~\cite{shlesinger-book,ary} for a number of health
conditions.

We do not find it inconceivable that persistence and repetition in
diverse self-regulating complex systems may emerge whenever memories
of recent events degrade preferentially to those of the distant past.
Consider, for example, the role of memory in health and
disease~\cite{maa0,maa1,alz-cullen}.  A frequent and burdensome
behavioral and psychological symptom of Alzheimer's disease and other
dementias involves persistent and repetitive
behavior~\cite{alz-cullen}.  Patients may hum a tune that never seems
to run out of verses, pose the same question dozens of times a day, or
pace the same stretch of floor for hours. They may also continuously
repeat words or phrases (i.e., echolalia).  Most importantly in the
context of our findings, patients that suffer from persistent and
continual repetition of questions (e.g., the inability to remember
directions) often show clear evidence of loss of recent memory and
immediate recall~\cite{alz-cullen}.  Memory loss frequently manifests
itself in early stages of the disease while repetitive actions are
less common in early dementia but increase in frequency in patients
with definite diagnoses of dementia~\cite{ready,boyd}.  While memory
impairment can clearly account for repetitive questioning, its role in
other repetitive actions (either as a cause or as a correlate) has
remained unclear thus far~\cite{alz-cullen}.  In view of these facts
and our reported findings, we find it plausible that recent memory
loss may bear a causal relationship with the repetitive behaviors seen
in Alzheimer's disease, in which memories of the distant past fade
last.

In summary, we have discovered a new mechanism underlying
trend-reinforcing dynamics: amnestic induction of persistence.
We have shown that loss of memory of the recent past
{can} cause persistence in otherwise nonpersistent non-Markovian
random walkers.  Whereas the loss of distant memories decreases
persistence~\cite{scv}, our results indicate that the loss of recent
memories actually increases persistence, allowing deviations from
Gaussian statistics.

\medskip
\medskip

\section*{Acknowledgements}
\medskip

We
acknowledge CNPq and FAPESP for financial assistance. GMV also thanks
FAPEAL.

\end{document}